\documentstyle[12pt]{article}

\catcode`\@=11

\@addtoreset{equation}{section}

\begin{document}

\baselineskip 24pt
\newcommand{\numero}{SWAT/16}
%Enter SHEP preprint number
\newcommand{\titre}{V,W AND X IN}
\newcommand{\titreb}{TECHNICOLOUR MODELS}
\newcommand{\auteura}{Nick Evans}
\newcommand{\addressa}{ }
\newcommand{\auteurc}{D.A. Ross }
\newcommand{\beq}{\begin{equation}}
\newcommand{\eeq}{\end{equation}}
\newcommand{\Fn}{\mbox{$F(p^2,\Sigma)$}}

\newcommand{\addressc}{Department Of Physics  \\   University   of
     Wales, Swansea\\ Singleton park, \\ Swansea \\ SA2 8PP \\ U.K. }
\newcommand{\abstrait}{
    Light techni-fermions and pseudo Goldstone bosons that contribute to
the electroweak radiative correction parameters V,W and X may relax the
constraints on technicolour models from the experimental values of the
parameters S and T. Order of magnitude estimates of the contributions to
V,W and X from light techni-leptons are made when the the techni-neutrino
has a small Dirac mass or a large Majorana mass. The contributions to
V,W and X from pseudo Goldstone bosons are calculated in a gauged chiral
Lagrangian. Estimates of V,W and X in one family technicolour models
suggest that the upper bounds on S and T should be relaxed by between
0.1 and 1 depending upon the precise particle spectrum.
      }

\begin{titlepage}
\hfill \numero
\vspace{.5in}
\begin{center}
{\large{\bf \titre }}
{\large{\bf \titreb}}
\bigskip \\by\bigskip\\ \auteura \bigskip \\ \addressc \\

\renewcommand{\thefootnote}{ }
\vspace{.9 in}
{\bf Abstract}
\end{center}
\abstrait
\bigskip \\
\end{titlepage}

\def\id{\rlap{1}\hspace{0.15em}1}

\section{Introduction}

     Precision  measurements  of   electroweak   parameters   are  now
sufficiently accurate to test  the  effects  of radiative corrections.
Heavy fermions do not decouple \cite{Lynn} from  these radiative effects
in broken
gauge theories and hence  these  measurements  provide  a probe of new
physics, such as technicolour \cite{TC},  at  mass  scales greater
than $M_Z/2$.
Peskin et al. \cite{Peskin} have developed an  elegant description of
the effects of
such new physics when the  new  particles  have  large masses of order
$1TeV$ in terms of the three parameters S,T and U

\beq \begin{array}{ccc}
\alpha S  &  =  &  4e^2  \frac{d}{dq^2}  [  \delta  \Pi_{33}  - \delta
\Pi_{3Q}]|_0 \\
&&\\
\alpha T & = &  \frac{e^2}{s_W^2c_W^2M_Z^2}  [ \delta \Pi_{11}(0) - \delta
\Pi_{33}(0)] \\
&&\\
\alpha U  &  =  &  4e^2  \frac{d}{dq^2}  [  \delta  \Pi_{11}  - \delta
\Pi_{33}]|_0 \end{array} \eeq

\noindent where:  $\delta  \Pi_{ab}$  are  the  contributions  to  the
standard electroweak theory gauge  bosons' vacuum polarizations, given
by $\Pi_{ab}^{\mu \nu}(q^2)  =  \Pi_{ab}(q^2)  g^{\mu  \nu} + (q^{\mu}
q^{\nu} {\rm terms}$), from the new  heavy  particles;  $s_W$ and $c_W$ are the
sine and cosine  of  the  weak  mixing  angle;  q  is  the gauge boson
momenta. A global fit to current  experimental data \cite{Peskin}
provides the upper
bounds $S \leq 0.5$ and $T \leq 0.6$ at the $95\%$ confidence limit. A
naive estimate of the contributions  to  S  and  T from a technicolour
model of electroweak symmetry breaking can  be  made by scaling up QCD
\cite{Peskin} or in  chiral  models  of  the  strong  interactions
\cite{NLCM}.  These custodial
$SU(2)$ preserving estimates give $S \sim  0.1N_{TC} / doublet$ and $T
\sim 0$ where $N_{TC}$ is  the  number  of  technicolours. All but the
most minimal technicolour models are ruled  out by this estimate of S.
 However, estimates  of  S  and T in one family
technicolour  models   with   custodial   isospin   breaking   in  the
technilepton sector, which manifests  as  either Dirac mass splittings
between the techni-electron and  techni-neutrino  \cite{appelandtern}
 or  as a Majorana mass
for the right handed techni-neutrino \cite{Ross},
 suggest  that models can be built
with both S and T $\leq 1$.

Recently the S,T and U formalism  has been extended to incorporate the
effects of new particles with  masses  ${\cal O}(M_Z)$ \cite{VWX}.
In the orginal formalism \cite{Peskin}
the new fermions were assumed  to  have large masses so that
their self energies could be described  well  by a Taylor expansion to
linear order: $\delta \Pi_{ab} \approx  A_{ab} + B_{ab}q^2$. Errors of
order $(M_Z/M_{new})^2$ were neglected.  Burgess  et  al. \cite{VWX}
 have shown that
this is equivalent to assuming that  three parameters vanish, V, W and
X

\beq \begin{array}{ccl}
\alpha  V  &   =   &    \delta   \Pi'_{ZZ}(M_Z^2)   -   \left[  \frac{
\delta\Pi_{ZZ}(M_Z^2) - \delta\Pi_{ZZ}(0)}{M_Z^2} \right] \\
&&\\
\alpha  W   &   =   &   \delta   \Pi'_{WW}(M_W^2)   -   \left[  \frac{
\delta\Pi_{WW}(M_W^2) - \delta\Pi_{WW}(0)}{M_W^2} \right] \\
&&\\
\alpha  X  &  =   &  -s_Wc_W\left[  \frac{\delta\Pi_{ZA}(M_Z^2)}{M_Z^2}  -
\Pi'_{ZA}(0) \right] \end{array} \eeq

\noindent where the prime  indicates  differentiation  with respect to
$q^2$. If these parameters are non-zero  then the S,T and U parameters
are given by the more general forms

\beq \begin{array}{ccl}
\alpha S  &  =  &  4  s_W^2c_W^2  \left[  \frac{  \delta\Pi_{ZZ}(M_Z^2)  -
\delta\Pi_{ZZ}(0)}{M_Z^2} \right] - 4s_Wc_W(c_W^2-s_W^2) \delta \Pi'_{ZA}(0) -
4s_W^2c_W^2 \delta \pi'_{AA}(0) \\
&&\\
\alpha  T  &  =   &   \frac{\delta   \Pi_{WW}}{M_W^2}  -  \frac{\delta
\Pi_{ZZ}(0)}{M_Z^2} \\
&&\\
\alpha  U  &  =   &   4s_W^2    \left[  \frac{  \delta\Pi_{WW}(M_W^2)  -
\delta\Pi_{WW}(0)}{M_W^2}    \right]  -4s_W^2c_W^2     \left[    \frac{
\delta\Pi_{ZZ}(M_Z^2) - \delta\Pi_{ZZ}(0)}{M_Z^2} \right] \\
&&\\
&& \hspace{1in} -4s_W^4\delta \Pi'_{AA}(0) -8s_W^3c_W \delta\Pi'_{ZA}(0)
\end{array} \eeq

A global fit to the  experimental  data  in  which all six parameters,
S,T,U,V,W and X, are  allowed  to  vary  simultaneously  gives the one
standard deviation bounds \cite{VWX2}

\beq \begin{array}{ccccccc}
S & \sim & -0.93 \pm 1.7 & \hspace{2cm} & V & \sim & 0.47 \pm 1.0 \\
&&\\
T & \sim & -0.67 \pm 0.92 & \hspace{2cm} & W & \sim & 1.2 \pm 7.0 \\
&&\\
U  & \sim & -0.60 \pm 1.1 & \hspace{2cm} & X & \sim & 0.1 \pm 0.58
\end{array} \eeq

The  inclusion  of  V,W  and  X  weakens  the  bounds  on  S,T  and  U
considerably. This analysis raises the possibility that a technicolour
model with new light particles with masses $\sim {\cal O}(M_Z)$ may be
experimentally viable. There are two  possible sources  of  such light
particles, light techni-fermions and
the pseudo-Goldstone bosons (PGB) that  occur  in  many technicolour models
with large global symmetries. An order of magnitude estimate of the
contributions to V,W, and X from light
techni-fermions can be obtained by calculating the one loop contribution
from a weakly interacting doublet \cite{VWX}. This naive estimate will be
sufficient to show whether V,W and X are large enough relative to S and T
to violate the assumption of the orginal global fit \cite{Peskin} that V,W
and X are all $\sim 0$. In section 2 of this paper
we review
this calculation
and extend it to find the contributions to V,W, and X from a techni-lepton
doublet
with a right handed Majorana mass. The  PGB  contributions  to S and T have
been calculated before \cite{Renken,appelandtern}. In section 3  we  review the
construction of a
Chiral Lagrangian of the interactions of the PGBs below the technicolour
confinement scale and present expressions for their contributions
to S,T,U,V,W, and X.
Finally we calculate
V,W and  X  in one  family  technicolour models with realistic  mass
spectra
(section 4).

\section{Light Techni-Fermions}

We consider a technicolour model with N left handed Weyl technifermions
$\psi_L^i$ and N right handed Weyl technifermions $\psi_R^i$
that transform either under a real
or complex representation of a technicolour group, {\cal G}, such as those
in ref \cite{ETCmodels}.
In addition to their
interactions under the technicolour group and the usual electroweak
interactions
the techni-fermions may have extended technicolour interactions from
physics above the scale $\Lambda_{TC}$ which can be represented by
4-Fermi operators. These
extended technicolour interactions may be strong at the technicolour
confinement scale \cite{Appel}. When the technicolour
group becomes strongly interacting at the scale $\Lambda_{TC}$ the
technifermions' approximate chiral symmetry is dynamically broken.
If the technifermions transform
under a complex representation of {\cal G} then the maximal global symmetry
breaking
pattern is given by

\beq SU(N)_L \otimes SU(N)_R \longrightarrow SU(N)_V \eeq

\noindent and if they transform under a real representaion of {\cal G} by

\beq SU(2N) \longrightarrow O(2N) \eeq

\noindent If the extended technicolour interactions are sufficiently strong
to distinguish different techni-fermions at the scale $\Lambda_{TC}$ then the
approximate global symmetries will be some sub-group of these groups.

The techni-fermions acquire dynamical masses of order $\Lambda_{TC}$; in
practice the precise values of the techni-fermion masses will be
dependent on the strengths of the extended technicolour interactions.
These techni-fermions can potentially be light (relative to $M_Z$) if
the scale $\Lambda_{TC}$ is low ($\sim M_Z$) as in, for example, one family
technicolour models with separate technicolour interactions on the
techni-quarks
and techni-leptons where one or other sector dominates electroweak symmetry
breaking, or models with other low scale, strongly interacting fermions
in addition to the electroweak symmetry breaking interactions \cite{Steve}.
The techni-fermions
must be integrated from the effective theory of the technicolour
interactions at current collider energies. We are interested in the
contribution to gauge boson self energies
(and hence V,W, and X) from strongly interacting loops of these
techni-fermions.
Such a non-perturbative calculation is beyond current technology especially
when the techni-fermions are light and custodial isospin is broken by the
extended technicolour interactions. We shall naively estimate the order of
magnitude of these contributions by calculating the one-loop perturbative
contributions to V,W and X of a weakly interacting fermion doublet with
momentum
independent mass.

We shall express our results in terms of the well known one-loop results
for an incoming gauge boson coupling to left handed fermions and an out
going gauge boson coupling to left and right handed fermions

\beq \begin{array}{ccc}
\Pi_{LL}(m_1,m_2,q^2) & = & -\frac{1}{4\pi^2} \int^1_0 dx \left( x(1-x)q^2 -
\frac{m^2}{2}
\right) \ln\left(
\frac{\Lambda^2}{m^2-x(1-x)q^2} \right) \\
&&\\
\Pi_{LR}(m_1,m_2,q^2) & = & -\frac{m_1m_2}{8\pi^2} \int^1_0 dx \ln \left(
\frac{\Lambda^2}{m^2-x(1-x)q^2} \right)  \end{array} \eeq

\noindent where $m^2 = xm_1^2 + (1-x)m_2^2$.

The contributions to the gauge boson self energies from a fermion doublet
, $\psi = (\psi_+,\psi_-)$, are thus given by the standard electroweak
couplings and we find

\begin{eqnarray}
\delta \Pi_{AA} & = & \sum_{a= \pm} 2N_{TC}e^2Q_a^2 \left[
\Pi_{LL}(m_a,m_a,q^2)
+ \Pi_{LR}(m_a,m_a,q^2) \right] \\
&& \nonumber \\
\delta\Pi_{ZA} & = & \sum_{a=\pm} \frac{e^2N_{TC}}{s_Wc_W}Q_a(T_{3a}-2s_W^2Q_a)
\left[ \Pi_{LL}(m_a,m_a,q^2) + \Pi_{LR}(m_a,m_a,q^2) \right] \\
&& \nonumber \\
\delta \Pi_{WW} & = & \frac{e^2N_{TC}}{2s_W^2} \Pi_{LL}(m_+,m_-,q^2) \\
&& \nonumber \\
\delta \Pi_{ZZ} & = & \sum_{a=\pm} \frac{e^2N_{TC}}{s_w^2c_W^2} \left[
(T_{3a}^2 -2s_W^2T_{3a}Q_a + 2s_w^4Q_a^2) \Pi_{LL}(m_a,m_a,q^2)  \right.\\
&& \hspace{2cm} \left.-
2s_W^2Q_a(T_{3a}-s_W^2Q_a) \Pi_{LR}(m_a,m_a,q^2) \right]
\end{eqnarray}

V,W and X are given in terms of these contributions to the gauge boson
self energies by Eqn(1.2).
As an example of the contributions to V,W and X from light techni-fermions
($m_{\psi} \sim M_Z/2$) we plot the contribution of a techni-lepton
doublet (we take $N_{TC}= 1$) with $m_{\nu} = 50 GeV$ in Fig 1. The maximal
values of V,W and X
result when both the techni-neutrino and the techni-electron are light
and are given by $V \sim -0.15N_{TC}$, $W \sim -0.1N_{TC}$ and $X \sim 0$

We may similarly estimate the contributions to V,W and X from a techni-lepton
doublet, $\Psi_L = (N,E)_L$,$E_R,N_R$ with a right handed Majorana neutrino
mass. Writing the  left and right handed degrees of freedom of N as two
Majorana  (self conjugate, $\psi^M=C(\bar{\psi}^M)^T$) fields $N_1^0$ and
$N_2^0$ all possible gauge invariant mass terms are then given by
\cite{majmass}

\beq {\cal L}_{mass} = -\frac{1}{2}(\bar{N}^0_1 \bar{N}^{0C}_2) \left(
\begin{array}{cc} 0 & m_D \\ m_D & -M \end{array} \right) \left( \begin{array}
{c} N_1^{0C} \\ N^0_2 \end{array} \right) \eeq

\noindent where $m_D$ is the Dirac mass and M is the Majorana mass. The
Majorana
mass eigenstate fields, $N_1$ and $N_2$ with masses $M_1$ and $M_2$ are given
by

\beq \left( \begin{array}{c} N_1^0 \\ N^{0C}_2 \end{array} \right) = \left(
\begin{array}{cc} ic_{\theta} \gamma_5 & s_{\theta} \\
-i s_{\theta} \gamma_5 & c_{\theta} \end{array} \right)  \left(
\begin{array}{c}
N_1 \\ N_2 \end{array} \right) \eeq

\noindent where

\beq c_{\theta}^2 = \frac{M_2}{M_1+M_2}, \hspace{2cm} s_{\theta}^2 = \frac{M_1}
{M_1+M_2} \eeq

\noindent and

\beq M=M_2-M_1, \hspace{2cm} m_D = \sqrt{M_1M_2} \eeq

\noindent and where we have made an axial $U(1)$ transformation on the $N_1$
field.
Using the self conjugacy properties of the Majorana fields we may rewrite the
charged and neutral weak currents in terms of the mass eigenstate fields
\cite{Gates} and obtain

\begin{eqnarray}
\bar{N}_L \gamma^{\mu} E_L & = & -ic_{\theta} \bar{N}_{1L} \gamma^{\mu} E_L +
s_{\theta} \bar{N}_{2L} \gamma^{\mu} E_L \\
&& \nonumber \\
\bar{N}_L \gamma^{\mu} N_L & = & -\frac{1}{2} c_{\theta}^2 \bar{N}_{1}
\gamma^{\mu}
\gamma_5 N_{L} - \frac{1}{2}s_{\theta}^2 \bar{N}_{2} \gamma^{\mu} \gamma_5 N_2
+ i s_{\theta}c_{\theta} \bar{N}_2 \gamma^{\mu} N_1 \end{eqnarray}

We thus obtain expressions for the gauge boson self energies
(remembering to include the additional
non-zero Wick contraction $<\psi^M \psi^M>$ \cite{Gates}). The neutrino only
contributes
to $\delta\Pi_{ZZ}$ and $\delta\Pi_{WW}$ and we find

\begin{eqnarray}
\delta \Pi_{WW} & = & \frac{e^2N_{TC}}{2s_W^2} \left[ c_{\theta}^2
\Pi_{LL}(M_1,m_e,q^2)
+ s_{\theta}^2 \Pi_{LR}(M_2,m_e,q^2) \right] \\
&& \nonumber \\
\delta \Pi_{ZZ} & = & \frac{e^2N_{TC}}{4s^2_Wc^2_W} \left[
c_{\theta}^4(\Pi_{LL}
(M_1,M_1,q^2) + \Pi_{LR}(M_1,M_1,q^2) ) \right. \nonumber \\
&& \hspace{2cm} + s_{\theta}^4 (\Pi_{LL}(M_2,M_2,q^2) -
\Pi_{LR}(M_2,M_2,q^2) ) \nonumber \\
&& \hspace{2cm} \left. + 2s_{\theta}^2c_{\theta}^2 (\Pi_{LL}(M_1,M_2,q^2) +
\Pi_{LR}(M_1,M_2,q^2) \right.\\
&& \hspace{2cm}  + (1-4s_W^2+8s^4_W)\Pi_{LL}(m_E,m_E,q^2) \nonumber \\
&& \hspace{2cm} \left. + (2s_W^2+4s_W^4)
\Pi_{LR}(m_E,m_E,q^2) \right] \nonumber
\end{eqnarray}

Only V and W, which are given in terms of Eqn(2.15) and Eqn(2.16)
 by Eqn(1.2), are changed
by the inclusion of the Majorana mass. As an example of the effects of
including the Majorana mass on S,T,V and W these quantities are plotted in
Fig 2 for a Dirac mass degenerate techni-lepton doublet ($m_D = 200 GeV, N_{TC}
=1$)
 for
varying Majorana mass. The regions of negative S and T have been reported
before \cite{gatesandtern}. Both V and W are slightly enhanced by the inclusion
of a Majorana mass. The maximal realistic values of V and W (corresponding
to a lightest mass eigenstate $>  M_Z/2$) are of order
$-0.04N_{TC}$ and $-0.01N_{TC}$ respectively.

\section{Pseudo Goldstone Bosons}

When the technicolour group becomes strongly interacting at the scale
$\Lambda_{TC}$ it breaks the approximate global symmetry of the techni-fermions
according to the symmetry breaking pattern in Eqn(2.1), Eqn(2.2) or some
appropriate sub-group of these patterns. There will be a Goldstone boson (GB)
associated with each of the broken generators. These bound states remain in
the effective theory below the scale at which the techni-fermions have
been integrated out. The low energy interactions
of these GBs may be described by a gauged Chiral Lagrangian \cite{Chadha}.
In the notation
of Peskin \cite{Chadha,PGBmass}  we write the techni-fermion fields as

\beq \Psi_L = \left( \begin{array}{c} \psi^i_L \\ \psi_R^{iC} \end{array}
\right), \hspace{2cm} \Psi_R = \Psi_L^C \eeq

\noindent where $\psi^C = -i\sigma_2 \psi^*$. The global symmetry
generators can then be written as $2N \times 2N$ matrices L acting on
$\Psi_L$ and R acting on $\Psi_R$ \footnote{We note that an $SU(N)_L$
global transformation on the techni-fermion fields will have both components
L and R since $\Psi_L$ and $\Psi_R$ both contain the left
handed degrees of freedom}. The construction

\beq U = {\rm exp} \left( \frac{2i\pi^a X^a}{f_{\pi}} \right) \eeq

\noindent where $\pi^a$ are the GBs, $X^a$ the $2N \times 2N$
broken generators of the global symmetry group of $\Psi_L$ and $\Psi_R$,
and $f_{\pi}$ the pion decay constant, transforms under the  global
symmetry transformations as

\beq U \longrightarrow LUR^{\dagger} \eeq

The effective theory of the GBs is then constructed from all chirally
invariant terms. At low energies we may perform an expansion in $q/f_{\pi}$
and keep only terms of lowest order in the covariant derivative. The first
non trivial term in the expansion is given by

\beq {\cal L}_{\chi} = \frac{f_{\pi}}{4} tr \left[ (D_{\mu}
U)^{\dagger} D^{\mu} U \right] \eeq

\noindent with

\beq D^{\mu}U = \partial^{\mu} U + ig_L A^{\mu}_a G^a_L U -ig_R A^{\mu}_a
U G^a_R \eeq

\noindent where we have written the gauge generators in terms of components
that act on $\Psi_L$, $G^a_L$, and those that act on $\Psi_R$, $G^a_R$. We
define $G_L^a$ and $G_R^a$ to contain the gauge couplings.

Gauge boson-PGB vertices may be obtained by explicitly expanding the
exponential; for example we obtain

\beq \begin{array}{ccc}
\Gamma_{\pi^a \pi^b A^{\mu c}} & = & i (k+q)^{\mu} tr \left\{ X^a [G_L^c,X^b] +
X^a
[G_R^c,X^b] \right\} \\
&&\\
\Gamma_{\pi^a \pi^b A^{\mu c} A^{\nu d}}& = &
tr \left\{ X^a [G_L^c,[G_R^d,X^b]] + X^a[G_R^d,[G_L^c,X^b]]
\right\} + c \leftrightarrow d \end{array} \eeq

In a realistic technicolour model the techni-fermions will decompose into
n $SU(2)_L$ doublets and 2n right handed singlets.
The standard model electroweak gauge generators may then be split into
components L and R

\beq \begin{array}{ccc} {\rm Gauge Boson} & L & R \\
&&\\
A^{\mu} & \left( \begin{array}{c|c} eQ & 0 \\ \hline 0 & -eQ \end{array}
\right) & \left( \begin{array}{c|c} -eQ & 0 \\ \hline 0 & eQ \end{array}
\right) \\
&&\\
Z^{\mu} & \frac{e}{s_W c_W} \left( \begin{array}{c|c}
\tau_3 \otimes \id_n - s_W^2Q & 0 \\ \hline 0 & s^2_WQ \end{array}
\right) & \frac{e}{s_W c_W} \left( \begin{array}{c|c}
-\tau^3 \otimes \id_n + s_W^2Q & 0 \\ \hline 0 & -s^2_WQ \end{array}
\right) \\
&&\\
W^{\mu}_{\pm} & g \left( \begin{array}{c|c} \tau^{\pm} \otimes \id_n & 0\\
\hline 0 & 0 \end{array} \right) & g \left( \begin{array}{c|c} -\tau^{\pm}
\otimes \id_n & 0 \\ \hline 0 & 0 \end{array} \right) \end{array} \eeq

\noindent where Q is the $2n \times 2n$ charge matrix of the Dirac
fermions $\psi$ and $\id_n$ is the $n \times n$ unit matrix.
Inserting these matrices
into Eqn(3.6) we obtain the vertices \cite{Chadha}

\beq \begin{array}{cccccc}
\Gamma_{\pi^a \pi^a W^{\pm \mu} W^{\pm \nu}} & = & 0 &
\Gamma_{\pi^{\pm} \pi^3 W^{\pm \mu}} & = &
\frac{ie}{2s_W} I_{3a} (p+p')^{\mu} \\
&&&&&\\
\Gamma_{\pi^a \pi^a Z^{\mu} Z^{\nu}} & = &
\frac{2e^2}{c_W^2}(Q^2_as_W^2 - Q_aI_{3a}) &
\Gamma_{\pi^a \pi^a Z^{\mu}} & = & \frac{ie}
{2s_Wc_W}(I_{3a}-2s^2_WQ_a)(p+p')^{\mu} \\
&&&&&\\
\Gamma_{\pi^a \pi^a A^{\mu} A^{\nu}} & = & 2Q^2_ae^2 &
\Gamma_{\pi^a \pi^a A^{\mu}} & = & ieQ_a(p+p')^{\mu} \\
&&&&&\\
\Gamma_{\pi^a \pi^a A^{\mu} Z^{\nu}} & = &
\frac{e^2}{s_Wc_W}(I_{3a}Q_a-2s^2Q_a^2) & & & \end{array} \eeq

\noindent where $I_{3}$ is the custodial SU(2) isospin quantum number of
the GB.

Three of the GBs associated with
the breaking of the gauged electroweak symmetry groups will be strictly
massless and are ``eaten" by the electroweak gauge bosons. These unphysical
states may be removed by working in Unitary gauge. The remaining
Goldstone modes only correspond to approximate global symmetries and hence
will acquire small masses ($\ll \Lambda_{TC}$) through their weak gauge
and extended technicolour interactions. Estimates of the masses of these
Pseudo Goldstone Bosons (PGB) have been made in one family technicolour models
(see for example Ref \cite{PGBmass}). We shall simply introduce explicit mass
terms
into the Lagrangian for these fields.

The PGBs contribute to gauge boson self energies through the loops
\cite{Renken,
Aoki}

\begin{eqnarray}
{\rm Fig 4a} & \sim & \frac{g_{\alpha \beta}}{16\pi^2} \left[ (C_{UV}+1)
(m_a^2+m_b^2-\frac{q^2}{3}) - 2{\cal F}(m_a,m_b,q^2) \right] \\
&& \nonumber \hspace{3cm} +
q_{\alpha} q_{\beta} \hspace{0.1cm}{\rm terms} \\
&& \nonumber \\
{\rm Fig 4b}  & \sim & \frac{g_{\alpha \beta}}{16\pi^2} \left[ {\cal
F}(m_a,m_a,0)
-(C_{UV}+1)m_a^2  \right] + q_{\alpha} q_{\beta} \hspace{0.1cm}
{\rm terms} \end{eqnarray}

\noindent where

\begin{eqnarray}
{\cal F}(m_a,m_b,q^2) & = & \int^1_0 dx \left[ m_a^2(1-x) +m_b^2x -q^2x(1-x)
\right] \nonumber \\
&& \hspace{3cm} \ln \left( \frac{m_a^2(1-x) +m_b^2x -q^2x(1-x)}{\Lambda^2}
\right) \\
&& \nonumber \\
C_{UV} & = & \frac{1}{\epsilon} - \gamma + 4\pi \end{eqnarray}

Loop diagrams involving higher numbers of GB fields are suppressed
at low energies by powers of $q/f_{\pi}$.

Using the vertices in Eqn(3.8) and the loop results in Eqn(3.9) and Eqn(3.10)
 we obtain expressions
for the gauge boson vacuum polarizations and hence V,W and X. For completeness
we also give expressions for S,T and U.

\begin{eqnarray}
S & = & \frac{1}{4\pi} \sum_a \left\{ \frac{(I_{3a}^2-4s^2_WI_{3a}Q_a)}{2}
\left[ \frac{{\cal F}(m_a,m_a,0) - {\cal F}(m_a,m_a,M_Z^2)}{M_Z^2}\right]
\nonumber \right. \\
&& \left.\hspace{3.5cm} + (2
s^4_WQ^2_a + (c_W^2-s_W^2)Q_aI_{3a}) {\cal F}'(m_a,m_a,0) \right\} \\
&& \nonumber \\
T & = & \frac{1}{8\pi s^2_Wc^2_WM_Z^2}\sum_{mult} \left( {\cal F}(m_+,
m_-,0) - \frac{1}{2}{\cal F}(m_+,m_3,0) - \frac{1}{2}{\cal F}(m_-,m_3,0)
\right)\\
&& \nonumber \\
U & = & \frac{1}{2\pi} \left\{
 \sum_{mult}   \left[ \frac{ {\cal F}
(m_+,m_3,0) + {\cal F}(m_-,m_3,0) - {\cal F}(m_+,m_3,M_W^2) - {\cal F}
(m_-,m_3,M_W^2)}{M_W^2} \right]\nonumber \right. \\
&& \hspace{.1cm} +
\sum_a (I_{3a}^2-4s_W^2I_{3a}Q_a+4s_W^4Q^2_a)
\left[ \frac{{\cal F}
(m_a,m_a,M_Z^2)-{\cal F}(m_a,m_a,0)}{M_Z^2} \right]
  \nonumber \\
&& \hspace{.1cm} \left. + \sum_a
4s^2_WQ_a(I_{3a}-s^2Q_{3a}) {\cal F}'(m_a,m_a,0) \right\}  \\
&& \nonumber \\
V & = & \frac{1}{8\pi s_W^2c^2_W} \sum_a (I_{3a}^2-
4s_W^2I_{3a}Q_a+4s^4_WQ_a^2) \left[ \frac{{\cal F}(m_a,m_a,M_Z^2) -
{\cal F}(m_a,m_a,0)}{M_Z^2}\right. \nonumber \\
&& \hspace{3.5cm}\left.- {\cal F}'(m_a,m_a,M_Z^2) \right] \\
&& \nonumber \\
W & = & \frac{1}{8\pi s^2} \sum_{mult}
\left[ \frac{{\cal F}(m_+,m_3,M_W^2) - {\cal F}
(m_+,m_3,0)}{M_W^2} -{\cal F}'(m_+,m_3,M_W^2)
\right. \nonumber \\
&& \hspace{3.5cm} \left. + \frac{{\cal F}(m_-,m_3,M_W^2)-
 {\cal F}(m_-,m_3,0)}{M_W^2} - {\cal F}'(m_-,m_3,M_W^2)
 \right] \\
&& \nonumber \\
X & = & \frac{1}{4\pi} \sum_a Q_a(I_{3a}-2s^2_WQ_a) \left[ \frac{ {\cal F}
(m_a,m_a,M_Z^2) - {\cal F}(m_a,m_a,0)}{M_Z^2} \right.\\
&& \hspace{3.5cm} \left. - {\cal F}'(m_a,m_a,M_Z^2)
\right] \end{eqnarray}

\noindent where the sum over ``a" and ``mult" are sums over the full
 PGB spectrum and
 custodial SU(2) $I = 1$ multiplets respectively.

In Fig 3 the contributions to V,W, and X are plotted for a mass degenerate
$I = 1$ PGB multiplet ($\pi^+,\pi^0,\pi^-$). The largest contributions
to V,W and X are obtained when the PGB masses are of order $M_Z/2$ and
are given by $V \sim -0.02$, $W \sim -0.05$ and $X\sim -0.02$.

\section{One Family Technicolour}

One family technicolour models \cite{ETCmodels} are particularly appealing
since they naturally give rise
to extended technicolour interactions that provide masses for the light
fermions.
These models with a full techni-family have 16 Weyl fermionic degrees
of freedom, $\Psi_L=(U^{\alpha}_L,D^{\alpha}_L,E_L,N_L$, $U^{\alpha C}_R$,
$D^{\alpha C}_R,E^C_R,N^C_R)$, and
are hence very tightly constrained by the orginal S and T analysis.
We shall consider the contributions to V,W and X in two one family
technicolour scenarios, when the techni-fermions transform under a complex
and under a real representation of the technicolour group.

\subsection{Complex Representation}

If the techni-fermions transform under a complex representation of the
technicolour group (eg an $SU(N)_{TC}$ group) the approximate global
symmetry breaking pattern is $SU(8)_L \otimes SU(8)_R
\rightarrow SU(8)_V$. The precise techni-fermion mass spectra will depend
upon perturbations to this symmetry breaking pattern from extended
technicolour interactions. We might expect the techni-fermion masses to
have the same characteristics as the light fermions intra-family spectra.
Thus we expect the techni-quarks to be substantially heavier than the
techni-leptons and the techni-neutrino to be very light. Mass splitting between
the techni-quarks is severly constrained by the experimental value of the
T parameter hence we shall assume that they are degenerate with $m_{t-q} >
300GeV$. The techni-quarks contribution to V,W and X will therefore be
effectively zero (the linear Taylor expansion of the contributions to
the $\Pi_{ab}$s
is a good approximation). Mass splitting between the techni-leptons is
much less well
constrained if both are light \cite{appelandtern}. As an example we take
$m_N = 50GeV$ and $m_E = 150 GeV$ and find

\beq V_{t-l} \sim -0.13N_{TC} \hspace{1cm} W_{t-l} \sim -0.01N_{TC}
\hspace{1cm} X_{t-l} \sim 0 \eeq

We note that the value of V drops by a factor of ten as we increase the
techni-neutrino mass to 100GeV.

In addition to the techni-fermions there are 63 (P)GBs associated
with the broken generators of $SU(8)_{axial}$

\beq X^a = \frac{1}{\sqrt{2}} \left( \begin{array}{c|c}  \Lambda^a & 0 \\
\hline
0 & -\Lambda^{aT} \end{array} \right) \eeq

\noindent where $\Lambda^a$ are the generators of SU(8). The GBs can be
classified by the generators $X^a$ and according to their electroweak and
colour quantum numbers

\beq \begin{array}{cccc}
\Theta^{\alpha} & \begin{array}{c} {\rm colour} \\ {\rm octets} \end{array} &
\left\{ \left( \begin{array}{c|c} \id_2 \otimes \lambda^{\alpha} & 0 \\
\hline
0&0  \end{array} \right), \left( \begin{array}{c|c} \tau^a \otimes
\lambda^{\alpha} & 0 \\
\hline
0 & 0 \end{array} \right) \right\} & m_{\Theta} \sim 250 GeV \\
&&&\\
T & \begin{array}{c} {\rm colour} \\ {\rm triplet} \end{array} &
\left\{ \left( \begin{array}{c|c} 0 & \Sigma \otimes \zeta \\
\hline \Sigma^T
\otimes \zeta^T & 0 \end{array} \right) \right\} & m_T \sim 150GeV \\
&&&\\
P & \begin{array}{c} {\rm colour} \\ {\rm singlets} \end{array} &
\left\{ \left( \begin{array}{c|c} \tau^a \otimes \id_3 & 0\\ \hline 0 & 0
\end{array} \right), \left(  \begin{array}{c|c} 0 & 0\\
\hline 0 & \tau^a
\end{array} \right), \left( \begin{array}{c|c} \frac{1}{3} \id_6 & 0 \\
\hline 0 & - \id_2 \end{array} \right) \right\} & m_P \sim 0 - 100 GeV
\end{array} \eeq

\noindent where $\tau^a$ are the generators of $SU(2)$, $\lambda^{\alpha}$
the generators of $SU(3)$, $\id_n$ the $n \times n$ unit matrix, $\zeta$
is a 3-vector of colour, and $\Sigma$ are the $2 \times 2$ matrices

\beq \Sigma = \left\{ \id_2,\tau^a,i\id_2,i\tau^a \right\} \eeq

The mass estimates are taken from reference \cite{PGBmass}. Amongst the $I=1$
Ps the PGBs with techni-quark constituents will have the largest couplings
to the electroweak gauge bosons ($F_{\pi}$) and hence will constitute the
majority of the admixture ``eaten" by the $W^{\mu}$ and $Z^{\mu}$ gauge
bosons. The remaining physical PGBs will be largely techni-lepton bound states
with light masses. To maximize the possible contributions to V,W and X we
take $m_{P} = 50GeV$

\beq V_{PGB_c} \sim -0.05 \hspace{1cm} W_{PGB_c} \sim -0.08 \hspace{1cm}
X_{PGB_c} \sim -0.01 \eeq

The contributions to V,W and X in one family technicolour models with
techni-fermions transforming under a complex representation of the technicolour
group are typically of order a few hundredths. If one or more techni-fermion
is very light ($< 100GeV$) then the value of V can rise to the order of
a few tenths.

\subsection{Real Representation}

If the techni-fermions transform under a real representation of the
technicolour group (eg an $SO(N)_{TC}$ group) then the global symmetry breaking
pattern is given by $SU(16) \rightarrow O(16)$. The gauged symmetries of
such a one family model do not forbid a Majorana mass for the right handed
techni-neutrino and hence this is an alternative explanation for the
proposed isospin splitting in the techni-lepton sector. The techni-lepton's
maximum contributions to V,W and X are then typically (for $m_D \sim
150-200GeV$ and $M \sim 400-700GeV$)

\beq V_{Maj} \sim -0.04N_{TC} \hspace{1cm} W_{Maj} \sim -0.01N_{TC}
\hspace{1cm} X_{Maj} \sim 0 \eeq

There are also 36 PGB anti-PGB pairs in addition to those discussed for
the complex representation
corresponding to the additional broken generators

\beq X^a = \left( \begin{array}{c|c} 0 & D \\ \hline
D^{\dagger} & 0 \end{array} \right) \eeq

\noindent Classifying these additional GBs by their electroweak and colour
quantum numbers we find

\beq \begin{array}{cccc}
DQ & \begin{array}{c} \rm{colour} \\ \rm{sextets} \end{array} &
\left\{ \left( \begin{array}{c|c} \underline{\sigma} \otimes \underline{s}
 & 0 \\
\hline 0 & 0 \end{array} \right) \right\} & m_{DQ} \sim 260GeV \\
&&&\\
LQ & \begin{array}{c} {\rm colour} \\ {\rm triplets} \end{array} &
\left\{ \left( \begin{array}{c|c} 0 & \underline{\sigma} \otimes \zeta \\
\hline \underline{\sigma} \otimes \zeta^{\dagger} & 0 \end{array} \right),
\left( \begin{array}{c|c} 0 & i\underline{\epsilon} \otimes \zeta
\\ \hline
-i\underline{\epsilon} \otimes \zeta^{\dagger} & 0 \end{array} \right),
\right.
 & m_{LQ} \sim 160GeV \\
& & \left. \left( \begin{array}{c|c} \underline{\epsilon} \otimes
\underline{a} & 0 \\
\hline 0 & 0 \end{array} \right) \right\} & \\
&&&\\
DL & \begin{array}{c} {\rm colour}\\ {\rm singlets} \end{array} &
\left\{ \left( \begin{array}{c|c} 0 & 0 \\ \hline
0 & \underline{\sigma} \end{array}
\right) \right\} & m_{DL} \sim 60 GeV  \end{array} \eeq

\noindent where $\underline{s}$ are the 6 components of the $3 \times 3$
symmetric tensor, $\underline{a}$ the 3 components of the $3 \times 3$
anitsymmetric tensor, $\underline{\sigma}$ the 3 components of the $2 \times 2$
symmetric tensor and $\underline{\epsilon}$ the $2 \times 2$ antisymmetric
tensor.

The total contributions to V,W and X from PGBs in real representation models
is thus

\beq V_{PGB_r} \sim -0.14 \hspace{1cm} W_{PGB_r} \sim -0.15 \hspace{1cm}
X_{PGB_r} \sim -0.05 \eeq

When the techni-fermions transform under a real representation of the
technicolour group one family technicolour models predict that V,W and X
are typically of order $-0.1$. Since the minimal one family $SO(N)_{TC}$
 technicolour model  that is
asymptotically free is SO(7) these values can be greatly enhanced by
a light techni-fermion. For example V can receive a contribution as large as
-0.3 if the techni-neutrino's light mass is a result of a right handed
Majorana mass or -0.8 if the techni-neutrino has a small Dirac mass.

\section{Conclusions}

The non-decoupling of heavy physics ($m_{new} \sim 1TeV$) beyond the
Standard Model of electroweak
interactions was orginally parameterized in terms of three variables
S,T and U \cite{Peskin}. A global fit to experimental data tightly
constrained technicolour models with large numbers of new strongly interacting
doublets such as one family technicolour models. Recently it has been pointed
out that the effects of new physics at mass scales close to the Z mass
could be included into the analysis through three additional variables
V,W and X. A global fit including these three new parameters places much
weaker bounds on the values of S, T and U. The parameters U and W only
enter into measurements made at the mass of the W boson and are hence less
well bounded than S,T,V and X.

In this paper we have naively estimated the contributions to V,W and X in
technicolour models with light techni-fermions by calculating the
contributions from a weakly interacting heavy fermion doublet. Whilst not
a quantitatively accurate calculation of V,W and X these results provide
an order of magnitude estimate of the contributions. These parameters
typically are of order a few hundredths unless the techni-fermions' masses
fall below 100GeV. In a one family technicolour model in which the
techni-neutrino's Dirac mass falls to 50GeV we find $V \sim -0.15N_{TC}$
which is comparable to S and T in models with realistic techni-fermion mass
spectra (typically S and T $< 1$). Estimates have also been made for
the contributions from a techni-lepton doublet with a light component
resulting from a see-saw mechanism in the techni-neutrino's mass matrix
due to a right handed Majorana mass. V and W are both enhanced in such
a scenario and for example we find $0>V \geq -0.04N_{TC}$.

Many technicolour models have large approximate chiral symmetries which
are broken by the strong technicolour interactions resulting in a plethora
of light pseudo Goldstone bosons ($m_{PGB} \ll \Lambda_{TC}$). The low energy
interactions of these PGBs have been described by a gauged chiral
Lagrangian and estimates made of their contributions to V,W and X. We
have estimated the contributions from PGB spectra in one family technicolour
models in which the techni-fermions transform under both real and complex
representations of the technicolour group. We conclude that both $V_{PGB}$
and $W_{PGB}$ lie between 0 and -0.15.

The variables S,T,U,V,W and X are normalized so that physical
observables are linear functions of the variables with coefficients of
the same order of magnitude \cite{VWX}.
Therefore, in the absence of
a global fit to the data in which V,W and X are allowed to vary over their
typical values in technicolour models, we can treat V,W and X as indicative
of the errors in the estimated values of S,T and U when compared with the
orginal global fit for these three variables alone. In the absence of light
techni-fermions the major contributions to V,W and X are from PGBs and
we find that the errors in the estimates of S and T are $\sim |V|$ (since
$X < V$ and W does not contribute to measurements on the Z mass) which
we estimate to be at most a few tenths. However, in one family technicolour
models or low scale technicolour models
\cite{Steve}  with light techni-fermions ($m_{t-f} < 100GeV$)
the contribution to V can rise
to a few tenths per technicolour. Thus, for example, in an $SO(7)_{TC}$
technicolour model the upper bounds on S and T must be relaxed by $\sim 1$.
We conclude that it is premature to rule technicolour models out based
solely on the global fit to S,T and U until up coming accelerators have
performed a search for new light techni-fermions and scalars. \vspace{1in}

\noindent {\bf Acknowledgments} The author would like to thank SERC for
in part supporting this work and Nick Dorey and Graham Shore for helpful
discussion.


\newpage

\begin{thebibliography}{99}
 \bibitem{Lynn} B W Lynn, M E Peskin and R G Stuart, in Physics at
      LEP, edited
      by J Ellis and R Peccei, CERN Report No. 86-02 1986; D C Kennedy
      and B W Lynn, Nucl Phys {\bf B322} (1989) 1.
 \bibitem{TC} E Farhi and L Susskind, Phys Report 74 No.3 (1981) 277.
 \bibitem{Peskin} M E Peskin and T Takeuchi, Phys Rev Lett {\bf65} (1990) 964;
      M E Peskin and T Takeuchi,  Phys  Rev  {\bf  D46} (1992) 381; M E
      Peskin, talk at the XXVI International Conference on High
      Energy Physics, Dallas, Texas, 1992.
 \bibitem{NLCM} B Holdom, J  Terning  and  K  Verbeek, Phys Lett
      {\bf B245} (1990) 612; {\bf B273} (1991) 549(E),
      J Terning, Phys Rev {\bf D44} (1991) 887,B Holdom,
      Phys Rev {\bf D45} (1992) 2534.
 \bibitem{appelandtern} T Appelquist and J Terning, Yale Preprint YCTP-P9-93.
 \bibitem{Ross} NJ Evans, SF King and DA Ross, Phys Lett {\bf B303} (1993)
      295; NJ Evans and DA Ross, Southampton University Preprint SHEP-92/93-23
 \bibitem{VWX} CP Burgess, I Maksymyk and D London, McGill Preprint 93/13.
 \bibitem{VWX2} CP Burgess, S Godfrey, H Konig, D London and I Maksymyk,
      McGill Preprint 93/24.
 \bibitem{Renken} R Renken and ME Peskin, Nucl Phys {\bf B211} (1983) 93.
 \bibitem{ETCmodels} E Fahri and L Susskind, Phys Rev {\bf D20} (1979) 3404;
      S Dimopolous, Nucl Phys {\bf B168} (1980) 69.
 \bibitem{Appel} T Appelquist and O Shapira, Phys Lett {\bf B249} (1990) 83.
 \bibitem{majmass} M Gell-Mann, P Ramond and R Slansky, in supergravity,
      ed van Nieuvenhuizen and D Freeman (North Holland, Amsterdam)(1979)
      315; T Yanagida, Prog Theor Phys {\bf B135} (1978) 66.
 \bibitem{Gates}EI Gates and KL Kowalski, Phys rev {\bf D37} (1987) 938.
 \bibitem{gatesandtern} E Gates  and  J  Terning,  Phys  Rev Lett {\bf67}
      (1991) 1840 ,S Bertolini and A   Sirlin,   Phys Lett {\bf B257}
      (1991) 179.
 \bibitem{Chadha} S Chadha and ME Peskin, Nucl Phys {\bf B185} (1981) 61.
 \bibitem{PGBmass} ME Peskin Nucl Phys {\bf B175} (1980) 197.
 \bibitem{Aoki} K Aoki et al, 1982 Supp Prog Theory Phys {\bf 73} 106.
 \bibitem{Steve} SF King, Phys Lett {\bf B314} (1993) 364.
\end{thebibliography}
\end{document}